\pdfoutput=1
\RequirePackage{fix-cm}
\documentclass[smallextended]{svjour3}       
\smartqed  

\usepackage{graphicx}

\newcommand{\panda}{${\sf \overline{P}ANDA}$\hspace*{1ex}}
\renewcommand{\textfloatsep}{0ex}

\begin{document}

\title{QCD studies and discoveries with $e^+$$e^-$ colliders\\ and future perspectives 
}



\author{Jens S\"oren Lange,\\ for the Belle Collaboration 
}


\institute{   Justus-Liebig-Universit\"at Giessen, II.~Physikalisches Institut\\
              Tel.: +49-641-9933241\\
              Fax: +49-641-9933239\\
              \email{soeren.lange@exp2.physik.uni-giessen.de}           
}

\date{Received: 08/30/2012}

\maketitle


\noindent
{\sf 
presented at\\
{\bf\sf SSP2012}\\
5$^{th}$ International Symposium on Symmetries in Subatomic Physics\\
2012/06/18-22, 2012, KVI Groningen, the Netherlands\\
}

\vspace*{-0.7cm}

\section{Introduction}

Charmonium- and bottomonium spectroscopy has been a flourishing 
field recently, as many new states were observed.
Among them, expected states (such as the $h_b$, $h_b'$, $\eta_b$, $\eta_b'$, 
described below) have been measured accurately and enable 
precision tests of QCD-based potential models
of the ansatz:

\vspace*{-0.3cm}
\begin{equation}
V(r) = \frac{-4}{3} \frac{\alpha_S}{r} + k r + V_{spin-orbit} + V_{spin-spin} + V_{tensor}
\label{ecornell}
\end{equation}

\vspace*{-0.1cm}
\noindent
The first two terms are a Coulomb-like term with the strong
coupling constant $\alpha_S$ and a linear term (phenomenologically
describing confinement) with a string constant $k$.
In addition, several non-expected states were found (such as
the X(3872) and Y(4260), described below), which do not fit into
any potential model. 
The results presented here were taken with the Belle \cite{nim_belle}
and BaBar experiments \cite{nim_babar}
in $e^+$$e^-$ collisions 
at beam energies 10.5-11.0~GeV (i.e.\ in the $\Upsilon$(nS) region). 
Charmonium-like states are e.g.\ produced in $B$ meson decays. 
Bottomonium-states are e.g.\ produced in radiative decays of 
$\Upsilon($n$S)$ resonances.

\vspace*{-0.3cm}

\section{Charmonium}

{\bf The X(3872) state} has been discovered in $B$ meson decay
in the decay X(3872) $\rightarrow$$J$/$\psi$$\pi^+$$\pi^-$
by Belle \cite{x3872_belle} and confirmed by other experiments 
\cite{x3872_babar} 
\cite{x3872_cdf2} 
\cite{x3872_d0}
\cite{x3872_lhcb}
\cite{x3872_cms}.
Among the recently newly observed and yet unexplained 
charmonium-like states (sometimes referred to as XYZ states) 
the X(3872) is the only one observed in several decay channels.
It has a surprisingly very narrow width $\Gamma_{X(3872)}$$\leq$1.2~MeV (90\% C.L.), 
although its mass is above the open charm threshold.

\begin{figure}[htb]
\includegraphics[width=\textwidth,height=3cm]{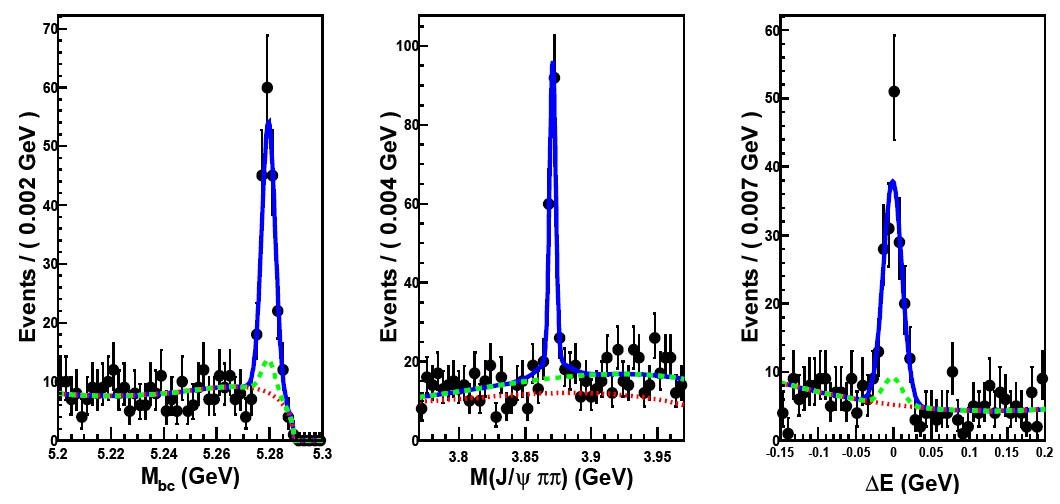}
\caption{
Invariant mass $m$($J$/$\psi$$\pi^+$$\pi^-$)
for $B^+$$\rightarrow$$K^+$X(3872)($\rightarrow$$J$/$\psi$$\pi^+$$\pi^-$).
}
\label{fx3872}
\end{figure}

\noindent
A recent mass measurement of the X(3872) at Belle was 
based upon the complete Belle data set of 711~fb$^{-1}$
(collected at the $\Upsilon$(4S) resonance), and is 
listed in Tab.~\ref{tx3872} in comparison with mass
measurements from other experiments. Fig.~\ref{fx3872}
shows the beam constrained mass $M_{\rm bc}$=$\sqrt{(E_{beam}^{cms}/2)^2-(p_B^{cms})^2}$ 
(with the energy in the center-of-mass system $E_{beam}^{cms}$ and the momentum
of the $B$ meson in the center-of-mass system $p_B^{cms}$), 
the invariant mass $m$($J$/$\psi$$\pi^+$$\pi^-$)
and the energy difference $\Delta$$E$=$E_B^{cms}$$-$$E_{beam}^{cms}$ 
(with the energy of the $B$ meson in the center-of-mass system $E^{cms}_B$).
Data and fit (as a result of a 3-dimensional fit to the observables 
shown) for the decay 
$B^+$$\rightarrow$$K^+$X(3872)($\rightarrow$$J$/$\psi$$\pi^+$$\pi^-$)
are shown (blue line: signal, dashed green line: background).
The fitted yield is 151$\pm$15 events. 
For details of the analysis procedure see \cite{x3872_belle_2011}.
As the X(3872) does not fit into any potential model prediction,
it was discussed as a possible S-wave $D^{*0}$$\overline{D}^0$
molecular state \cite{tornqvist_1} \cite{tornqvist_2}. 
In this case, the binding energy $E_b$ would be given 
by the mass difference $m$(X)$-$$m$($D^{*0}$)$-$$m$($D^0$). 
Including the new Belle result, the new world average mass of the X(3872)
is $m$=3871.68$\pm$0.17~MeV \cite{pdg}.
Using the current sum of the masses $m$($D^0$)+$m$($D^{*0}$) =3871.84$\pm$0.28~MeV \cite{pdg},
a binding energy of \mbox{$E_b$=$-$0.16$\pm$0.33~MeV} can be calculated,
which is surprisingly small.
As $E_b$ is inverse proportional to the squared scattering length $a$,
and the radius can in first order be approximated by $<$$r$$>$=$a$/2 \cite{braaten},
this would indicate a very large radius of the molecular state {\cal O}($\geq$10~fm).

\begin{table}[tb]
\caption{Mass measurements of the X(3872).}
\label{tx3872}
\begin{tabular}{|l|l|l|}  
\hline
Experiment & Mass of X(3872) & \\
\hline
\hline
CDF2 & 3871.61$\pm$0.16$\pm$0.19~MeV & \cite{x3872_cdf2} \\
\hline
BaBar ($B^+$) & 3871.4$\pm$0.6$\pm$0.1~MeV & \cite{x3872_babar} \\
\hline
BaBar ($B^0$) & 3868.7$\pm$1.5$\pm$0.4~MeV & \cite{x3872_babar} \\
\hline
D0 & 3871.8$\pm$3.1$\pm$3.0~MeV & \cite{x3872_d0} \\
\hline
Belle & 3871.84$\pm$0.27$\pm$0.19~MeV & \cite{x3872_belle_2011} \\
\hline
LHCb & 3871.95$\pm$0.48$\pm$0.12~MeV & \cite{x3872_lhcb} \\
\hline
\hline
New World Average & 3871.68$\pm$0.17~MeV & \cite{pdg} \\
\hline
\end{tabular}
\end{table}

\noindent
One of the surprising properties of the X(3872) is isospin violation.
It was found, that in the decay X(3872)$\rightarrow$$J$/$\psi$$\pi^+$$\pi^−$ 
the invariant mass peaks at the mass of the $\rho^0$ meson. 
The $\rho^0$ carries isospin $I$=0, but the initial state (if assumed 
to be a pure $c$$\overline{c}$ state) has $I$=0 (as it would not contain
any $u$ or $d$ valence quarks). 
One of the possible explanations might be $\rho$/$\omega$ mixing \cite{terasaki}.
There are only two additional isospin violating transitions known 
in the charmonium system \cite{pdg}, namely
$\psi'$$\rightarrow$$J$/$\psi$$\pi^0$ 
(${\cal B}$=1.3$\pm$0.1$\cdot$10$^{-3}$)
and $\psi'$$\rightarrow$$h_c$$\pi^0$ 
(${\cal B}$=8.4$\pm$1.6$\cdot$10$^{-4}$).
For the X(3872) the branching fraction of isospin violating 
transistion is (among the known decays) 
order of {\cal O}(10\%) and thus seems to be largely enhanced.

\begin{figure}[htb]
\centerline{\includegraphics[width=1.0\textwidth]{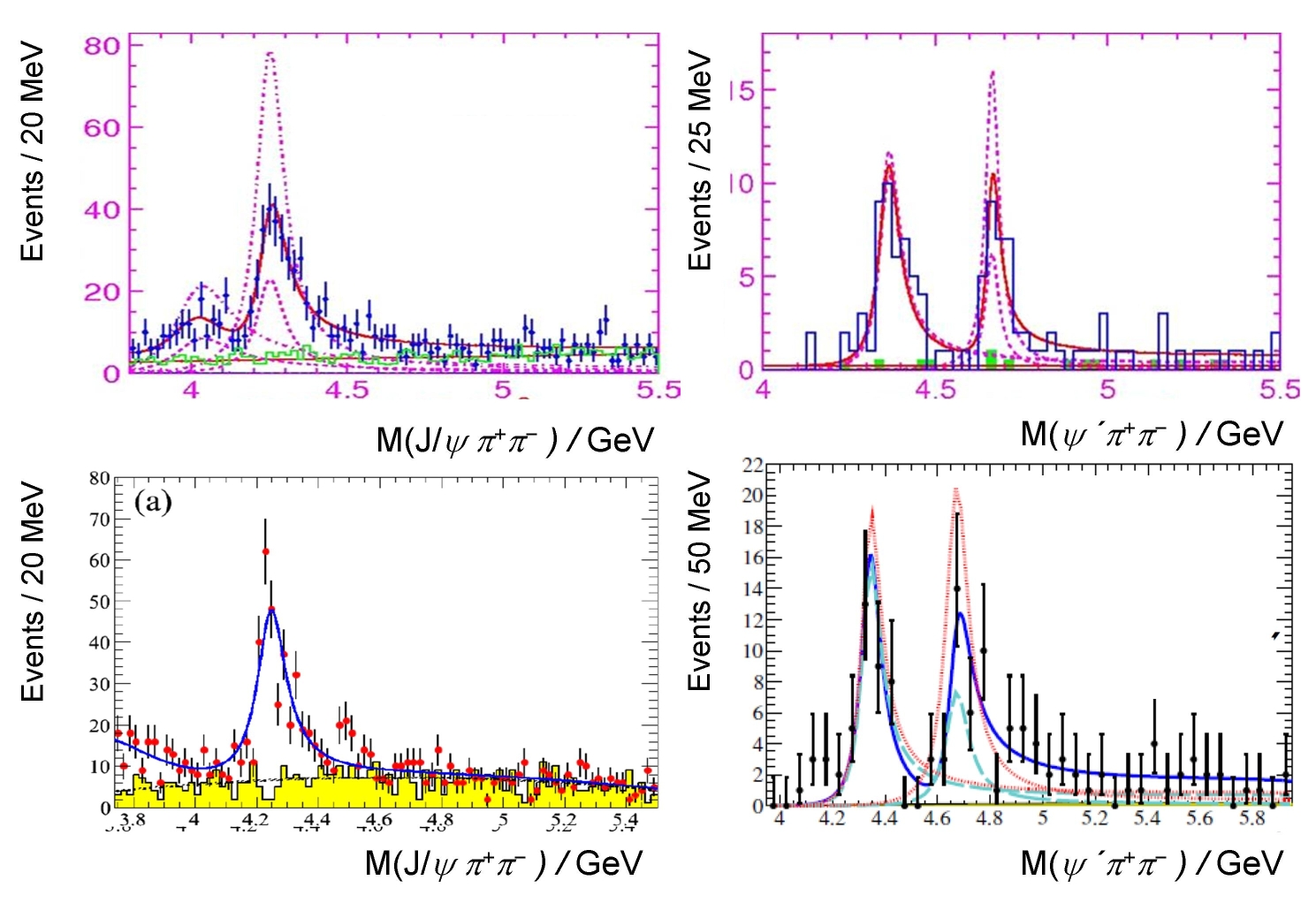}}
\caption{Observations of the Y states. 
Invariant mass $m$($J$/$\psi$$\pi^+$$\pi^-$) 
at Belle \cite{y4260_belle_2} (top left) 
and at BaBar \cite{y4260_babar_2} (bottom left).
Invariant mass $m$($\psi'$$\pi^+$$\pi^-$) 
at Belle \cite{y4350_belle} (top right) and 
at BaBar \cite{y4350_babar}.\label{fy4260}}
\end{figure}

\begin{figure}[htb]
\centerline{\includegraphics[width=0.6\textwidth]{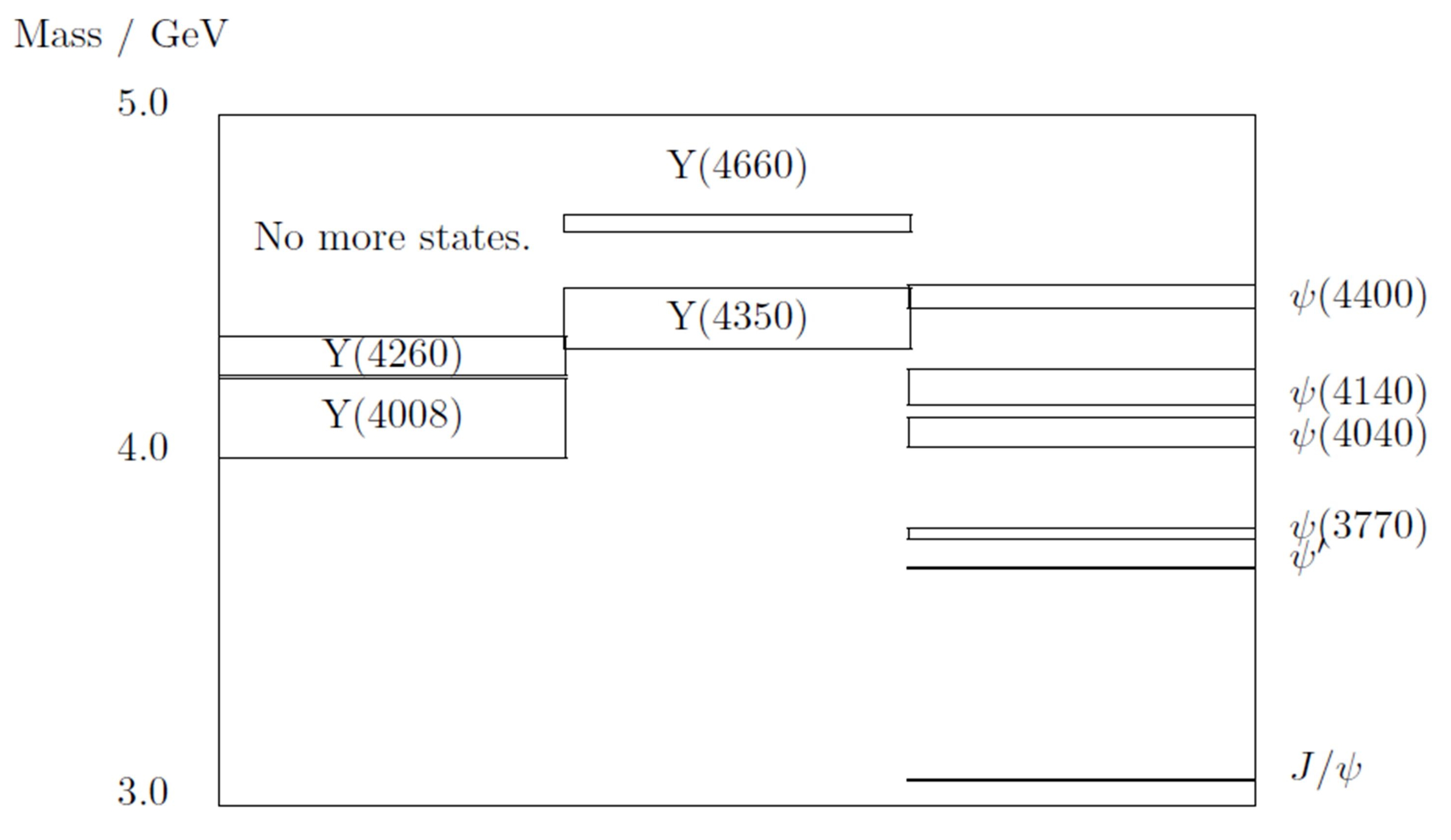}}
\caption{Level scheme for J$^{PC}$=1$^{--}$ states:
states decaying into $J$/$\psi$$\pi^+$$\pi^-$ (left column),
states decaying into $\psi'$$\pi^+$$\pi^-$ (center column),
and known $\psi$ states (radial quantum number $n$=1,...,6).\label{fy4260_scheme}}
\end{figure}

\noindent
{\bf The Y(4260) family}. Another new charmonium-like state was observed by BaBar
and confirmed by several experiments (see Tab.~\ref{ty4260}
for a list of the measured masses and widths) 
at a high mass of $m$$\simeq$4260~MeV,
far above the $D$$\overline{D}$ threshold. The width
is $\leq$100~MeV, which is quite narrow for such a 
high state. The observed decay is again a $\pi^+$$\pi^-$
transition to the $J$/$\psi$, similar to the first observed
decay of the X(3872). However, the production mechanism 
is not $B$ meson decay but instead ISR (initial state radiation), 
i.e.\ $e^+$$e^-$$\rightarrow$$\gamma$Y(4260), 
i.e.\ a photon is radiated by either the $e^+$ or the $e^-$ in the initial
state, lowering the $\sqrt{s}$ and producing the Y(4260) 
by a virtual photon. 
In fact, not only one state, but four states have been observed and are shown
in Fig.~\ref{fy4260}, i.e.\ the Y(4008), the Y(4260), the Y(4250) and the Y(4660).
In a search by Belle no additional state up to $m$$\leq$7~GeV was found.
All the Y states must have the quantum numbers
$J^{PC}$=$1^{--}$, due to the observation in an initial state radiation process.
As an intriguing fact, there are known and assigned 
$J^P$=$1^{--}$ charmonium states:
$J$/$\psi$, $\psi(2S)$, $\psi$(4040), $\psi$(4160) and $\psi$(4415).
Thus, there is a clear over-population of 1$^{--}$ states 
in the $m$$\geq$4~GeV region. 
Although they partially even overlap with their widths, 
apparently there seems to be no mixing: {\it (a)} no mixing among them, 
i.e.\ the Y(4008) and the Y(4260) decay to $J$/$\psi$$\pi^+$$\pi^-$, 
and the Y(4350) and the Y(4660) decay to $\psi'$$\pi^+$$\pi^-$, 
and neither of one has been observed in the other channel, and 
{\it (b)} no mixing with $\psi$ states with the Y states was observed
so far. 
The pattern of the Y states appears non-trivial (see Fig.~\ref{fy4260_scheme}): 
two non-mixing doublets without parity flip and without charge flip. 
It remains completely unclear what the underlying symmetry is.
In addition, there is no obvious pattern so far, 
how the masses of the $\psi$ states 
and the masses of the Y states might be related.

\begin{table}
\caption{Summary of the mass and width measurements of the Y(4260).\label{ty4260}}
\begin{tabular}{|l|l|l|l|l|l|l|}
\hline
 & 
BaBar \cite{y4260_babar_1} & 
CLEO-c \cite{y4260_cleo-c} & 
Belle \cite{y4260_belle_1} & 
Belle \cite{y4260_belle_2}  &
BaBar \cite{y4260_babar_2} &
BaBar \cite{y4260_babar_3} \\
\hline
${\cal L}$ &
211~fb$^{-1}$ &
13.3~fb$^{-1}$ &
553~fb$^{-1}$ &
548~fb$^{-1}$ &
454~fb$^{-1}$ &
454~fb$^{-1}$ \\
\hline
N & 
125$\pm$23 & 
14.1$^{+5.2}_{-4.2}$ & 
165$\pm$24 & 
324$\pm$21 &
344$\pm$39 &
$-$ \\
\hline
Significance & 
$\simeq$8$\sigma$ & 
$\simeq$4.9$\sigma$ & 
$\geq$7$\sigma$ & 
$\geq$15$\sigma$ &
$-$ &
$-$\\ 
\hline
$m$ / MeV & 
4259$\pm$8$^{+2}_{-6}$ & 
4283$^{+17}_{-16}$$\pm$4 & 
4295$\pm$10$^{+10}_{-3}$ & 
4247$\pm$12$^{+17}_{-32}$ &
4252$\pm$6$^{+2}_{-3}$ &
4244$\pm$5$\pm$4 \\
\hline
$\Gamma$ / MeV & 
88$\pm$23$^{+6}_{-4}$ & 
70$^{+40}_{-25}$ & 
133$\pm$26$^{+13}_{-6}$ & 
108$\pm$19$\pm$10 &
105$\pm$18$^{+4}_{-6}$ &
114$^{+16}_{-15}$$\pm$7 \\
\hline
\end{tabular}
\end{table}


\vspace*{-0.7cm}

\section{Bottomonium}

\begin{figure}[htb]
\centerline{\includegraphics[width=0.8\textwidth]{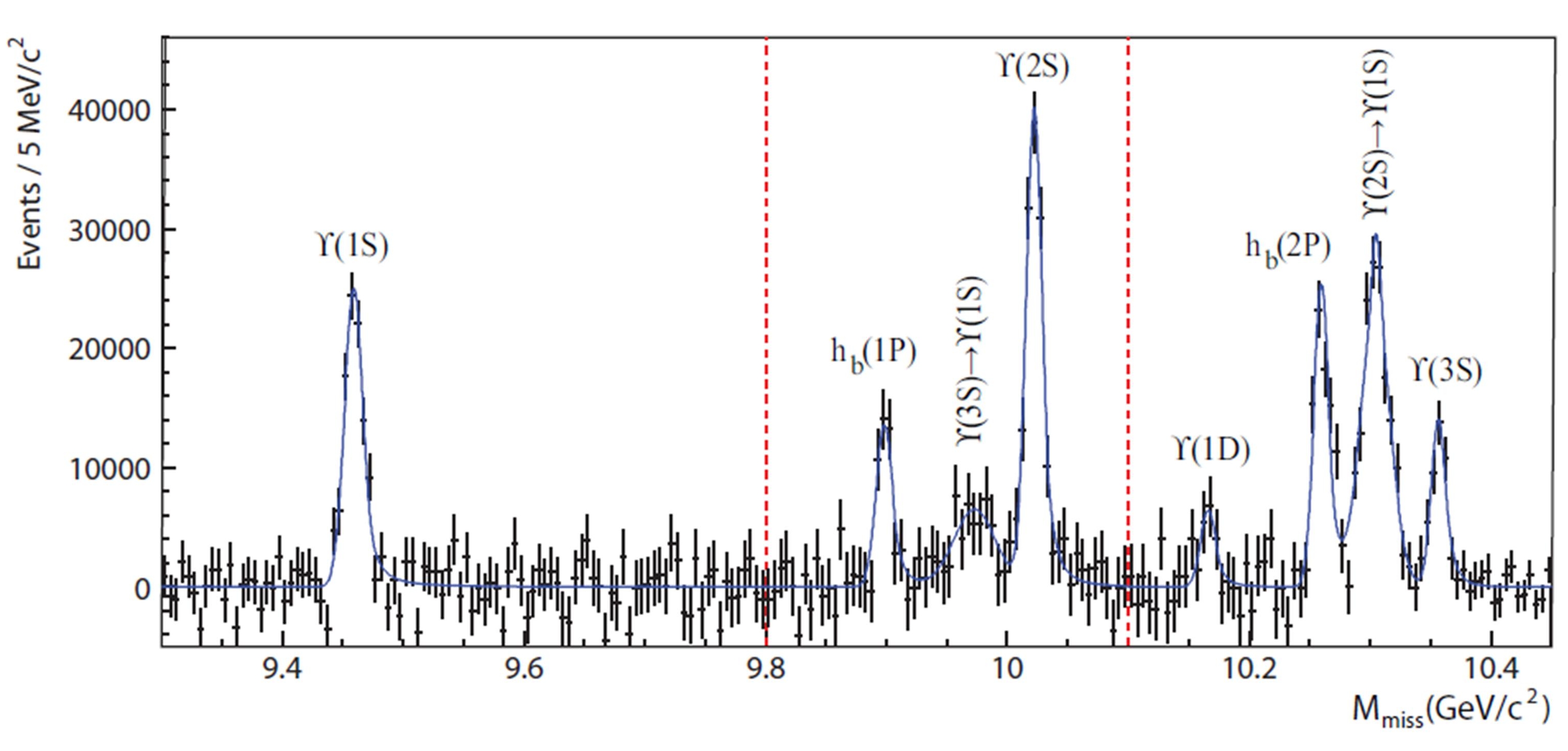}}
\caption{Observation of the $h_b$(1S) and $h_b$(2S) at Belle. 
For details see text.\label{fhb}}
\end{figure}

\noindent
{\bf The $h_b$(1P) and $h_b$(2P) states.} In a recent analysis by Belle, by usage of 
a particular technique, namely study of missing masses to
a $\pi^+$$\pi^-$ pairs in $\Upsilon$(5S) decays \cite{hb_belle}.
Fig.~\ref{fhb} shows the background-subtracted 
missing mass for a $\Upsilon$(5S) data set of 121.4~fb$^{-1}$. 
Among several known states such as the $\Upsilon$(1S), $\Upsilon$(2S), 
$\Upsilon$(3S) and $\Upsilon$(1D), 
there are addititional peaks arising from the transistions 
$\Upsilon$(3S)$\rightarrow$$\Upsilon$(1S)$\pi^+$$\pi^-$
$\Upsilon$(2S)$\rightarrow$$\Upsilon$(1S)$\pi^+$$\pi^-$, 
with the $\Upsilon$(3S) and $\Upsilon$(2S) 
being produced in the decay of the primary $\Upsilon$(5S). 
In addition to the expected signals, first observations 
of the bottomonium singlet 
P-wave states $h_b$(1P) and $h_b$(2P) were made. 
Their measured masses are 
$m$=9898.3$\pm$1.1$^{+1.0}$$_{-1.1}$~MeV and 
$m$=10259.8$\pm$0.6$^{+1.4}$$_{-1.0}$~MeV, respectively.  
For the $h_b$, this measurement is consistent with the first evidence
(3.1$\sigma$ stat.\ significance) by BaBar in $\Upsilon$(3S) decays 
with a mass of 9902$\pm$4(stat.)$\pm$2(syst.)~MeV \cite{hb_babar_pi0}.
The masses can be compared to predictions from 
potential model calculations \cite{potential_bb}
with 9901 MeV and 10261 MeV, respectively, 
i.e.\ the deviations are only 2.7 MeV and 1.2~MeV.


\begin{figure}[htb]
\centerline{\includegraphics[width=0.8\textwidth]{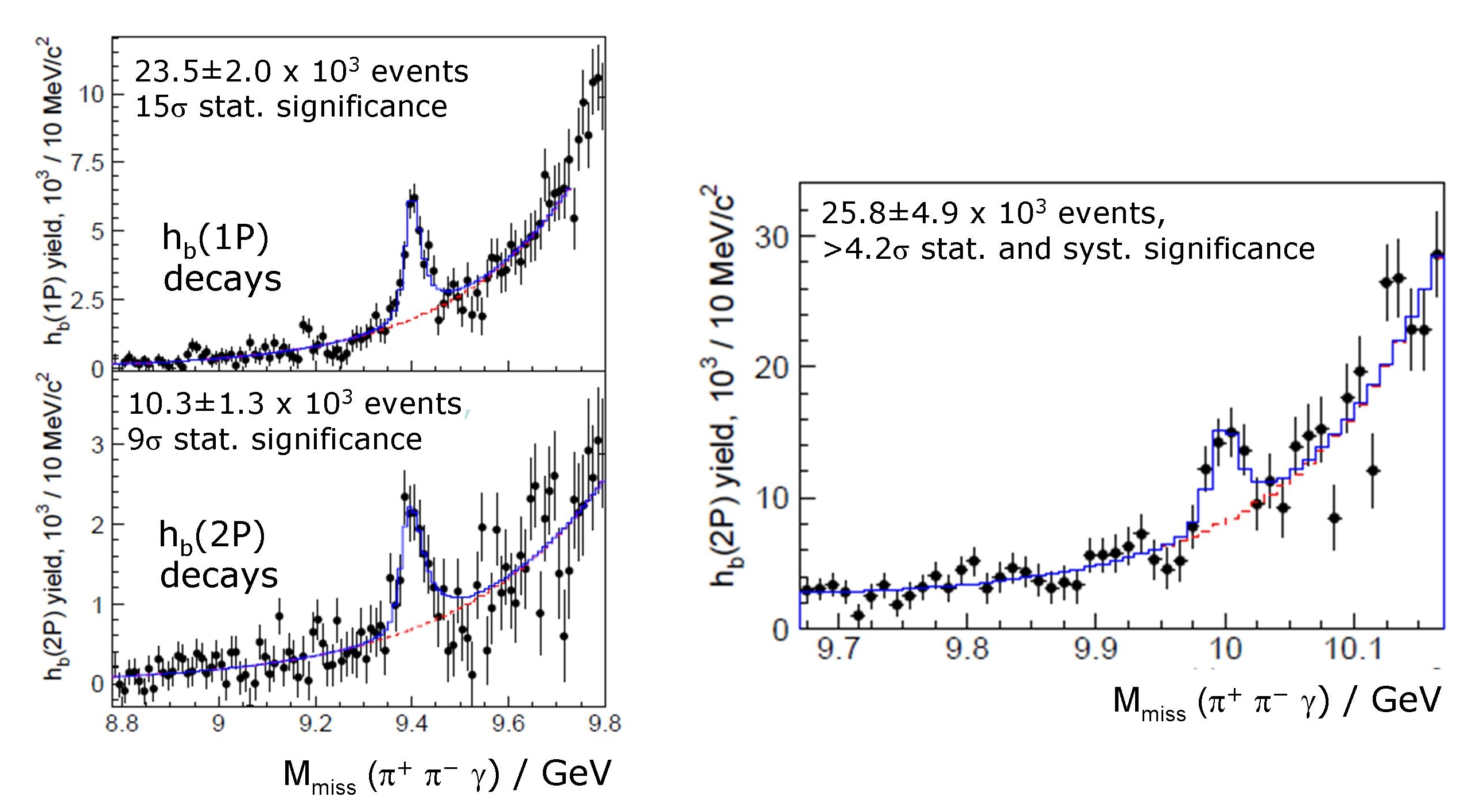}}
\caption{Observations of the $\eta_b$(1P) (left) and $\eta_b$(2P) (right) 
at Belle. For details see text.\label{fetab}}
\end{figure}

\noindent
{\bf The $\eta_b$(1S) and $\eta_b$(2S) states.}
The $\eta_b$(1S) is the bottomonium ground state 
$1^1S_0$ with J$^{PC}$=0$^{-+}$.
It was discovered by BaBar in the radiative decay 
$\Upsilon$(3S)$\rightarrow$$\gamma$$\eta_b$.
The measured mass was 9388.9$^{+3.1}_{-2.3}$(stat)$\pm$2.7(syst)~MeV, 
The observation was confirmed by CLEO III using 6 million Upsilon(3S) decays
with a measured mass $m$=9391.8$\pm$6.6$\pm$2.0~MeV.
The observation of the $h_b$ (see above) by Belle also enabled a search
for the radiative decay $h_b$(1P)$\rightarrow$$\eta_b$(1S)$\gamma$, 
which was observed with a very high significance $>$13$\sigma$
in a dataset of 133.4~fb$^{-1}$ at the $\Upsilon$(5S) and in the 
nearby continuum \cite{etab_belle}. In addition, even the $\eta_b$(2S)
was observed in $h_b$(2P)$\rightarrow$$\eta_b$(2S)$\gamma$.
Fig.~\ref{fetab} shows the 
$\pi^+$$\pi^-$$\gamma$ missing mass for the case of the $\eta_b$(1S) (left)
and $\eta_b$(2S) (right), 
where the charged pion pair originates from the transition
$\Upsilon$(5S)$\rightarrow$$h_b$(1P,2P)$\pi^+$$\pi^-$.
The measured masses are $m$($\eta_b$(1S))=9402.4$\pm$1.5$\pm$1.8~MeV
and $m$($\eta_b$(2S))=9999.0 $\pm$3.5$^{+2.8}_{-1.9}$~MeV.
Due to the high resolution, this measurement also enabled
the measurement of the width of the $\eta_b$ as 
$\Gamma$=10.8$^{+4.0}_{-3.7}$$^{+4.5}_{-2.0}$,
which is consistent with the expectation from 
potential models to 5$\leq$$\Gamma$$\leq$20~MeV.
The measurements of the $\eta_b(1S)$ and $\eta_b(2S)$ allow
precision determination of the hyperfine mass splittings 
$\Upsilon$(1S)-$\eta_b$(1S) and $\Upsilon$(2S)-$\eta_b$(2S),
using the masses of the $\Upsilon$(1S) and $\Upsilon$(2S)
from \cite{pdg}. The mass splittings are listed in Tab.~\ref{thfsplit}.
The splittings are in good agreement 
with the expectation from a potential model with relativistic 
corrections \cite{potential_bb}.
and Lattice QCD calculations 
with kinetic terms up to {\cal O}($v^6$) 
\cite{lattice_bb_new_meinel}. 
However, lattice QCD calculations to {\cal O}($v^4$) with charm sea
quarks predict higher splittings which are $\simeq$10~MeV larger. 
Note that perturbative non-relativistic QCD calculations
up to order ($m_b$$\alpha_S$)$^5$ predict significant 
smaller splittings e.g.\ 39$\pm$11$^{+9}_{-8}$~MeV
\cite{pNRQCD_bb}. 

\begin{table}
\begin{tabular}{|l|l|l|l|l|}
\hline
&  Belle \cite{etab_belle} & Potential \cite{potential_bb} & LQCD \cite{lattice_bb_new_hpqcd}) & LQCD \cite{lattice_bb_new_meinel}\\
\hline
$\Upsilon(1S)$$-$$\eta_b$ & 57.9$\pm$2.3 MeV & 60.0 & 70$\pm$9 MeV & 60.3$\pm$5.5$\pm$5.0$\pm$2.1~MeV\\
\hline
$\Upsilon(2S)$$-$$\eta_b'$ & $24.3^{+4.0}_{-4.5}$ MeV & 30.0 & 35$\pm$3 MeV & 23.5$\pm$4,1$\pm$2.1$\pm$0.8~MeV\\
\hline
\end{tabular}
\caption{Bottomonium hyperfine splittings: measurement, 
calculated by potential model and calculated by Lattice QCD (LQCD).\label{thfsplit}}
\end{table}

\begin{figure}[htb]
\centerline{\includegraphics[width=0.8\textwidth]{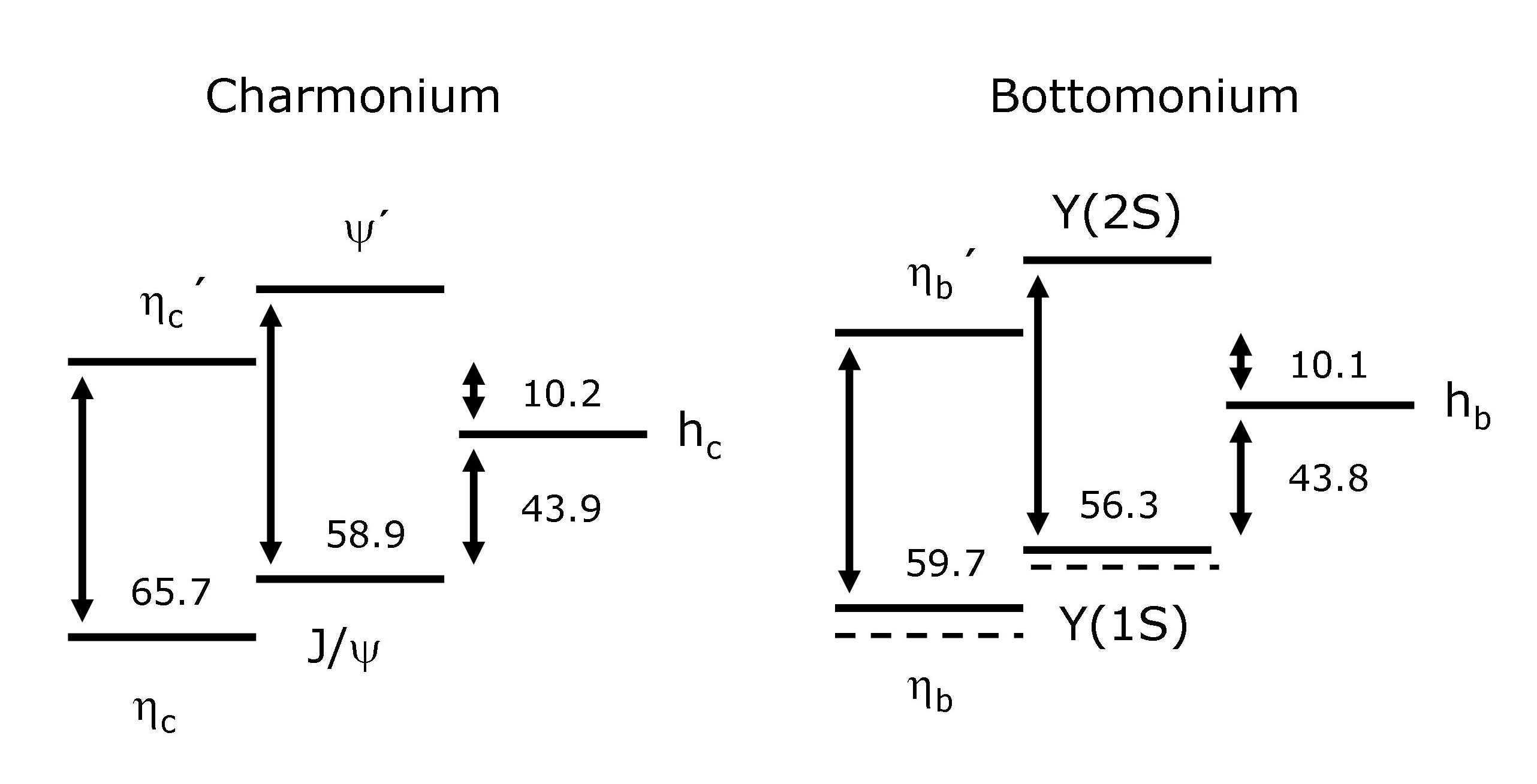}}
\caption{Mass splittings (in MeV) based upon the new measurements \cite{etab_belle} 
of the $h_b$, $\eta_b$ and $\eta_b'$, 
using masses from \cite{pdg} for the other states, for charmonium (left) and bottomonium (right).
The dotted lines indicate levels for the theoretical case 
of exact flavour independance.\label{fcc_bb_quigg_test}}
\end{figure}

\noindent
The new mass measurements enable for the first time a precision test 
of the flavour independance of the $c$$\overline{c}$ and $b$$\overline{b}$ 
systems. The important question is, if the level spacing is independant from the 
quark mass. According to \cite{cc_bb_quigg_test}, for a potential of
the form $V$($r$)=$\lambda$$r^\nu$ the level spacing is 
$\Delta$$E$$\propto$(2$\mu$/$\hbar^2$)$^{-\nu/(2+\nu)}$$|$$\lambda$$|$$^{2/(2+\nu)}$.
where $\mu$ is the (reduced) quark mass. For a pure Coulomb potential
($\nu$=$-$1), which should be dominating for the low lying states, 
this leads to $\Delta$$E$$\propto$$\mu$, This would imply that 
the level spacing would increase linearly with mass, i.e.\
$\Delta$$E$($b$$\overline{b}$)$\simeq$3$\Delta$$E$($c$$\overline{c}$).
For a pure linear potential it would be 
$\Delta$$E$$\propto$$\mu^{-1/3}$, thus the level spacing would 
decrease for higher quark masses, i.e.\ 
$\Delta$$E$($b$$\overline{b}$)$\simeq$0.5$\Delta$$E$($c$$\overline{c}$).
As can be seen in Fig.~\ref{fcc_bb_quigg_test}, 
for the mass splittings involving the $h_b$ (S=0, L=1) 
the agreement between
$c$$\overline{c}$ and $b$$\overline{b}$ is excellent,
i.e.\ 10.2 vs.\ 10.1~MeV and 43.9 vs.\ 43.8~MeV.
There are two possible explanations of this
remarkable symmetry:\\
{\it (1)} For a pure logarithmic potential V(r)=$\lambda$ln$r$ (i.e.\ the limit
$\nu$$\rightarrow$0) the level spacing is 
$\Delta$$E$$\propto$$\lambda$$\mu^0$. This means, 
the flavour independance would be 
strictly fulfilled. However, due to the vector nature of the gluon
the short-range potential must have a Coulomb-like part, and a 
pure logarithmic potential is therefore not possible.\\
{\it (2)} The other way to reach the flavour independance is,
that Coulomb potential 
($\Delta$$E$($b$$\overline{b}$)$\simeq$3$\Delta$$E$($c$$\overline{c}$)
(see above) and the linear potential 
$\Delta$$E$($b$$\overline{b}$)$\simeq$0.5$\Delta$$E$($c$$\overline{c}$)
(see above) cancel each other quantitatively and exact. 
It also implies that the size of the according $\lambda$ pre-factors 
($\lambda$=$-$4/3$\alpha_S$ for the Coulomb-like potential and 
$\lambda$=$k$ for the linear potential) just seem to have
the exactly correct size assigned by nature in a fundamental way.\\
For the ground states (S=0, L=0) the agreement
of the mass splittings between $c$$\overline{c}$ and $b$$\overline{b}$
is not as good, i.e.\ 65.7 vs.\ 59.7~MeV, and may point to the fact, that
there is an additional effect which lowers the $\eta_c$ mass. 
This might be mixing of the $\eta_c$ with the light quark states 
of the same quantum number 0$^{-+}$ (i.e.\ $\eta$ or $\eta'$). 

\vspace*{-0.7cm}

\section{Future Projects}

One of the important steps would be to measure not only the {\it masses}
of newly observed states, but also the {\it widths}. 
As many states have natural widths in the sub-MeV regime, 
the experiments must be able to reach according precision.
Two future projects should be mentioned here.\\ 
The Belle II experiment \cite{belle2_tdr} is an upgrade of the Belle experiment 
with a luminosity increased by a factor 
$\leq$40 to {\cal L}=8$\cdot$10$^{35}$~cm$^{-2}$~s$^{-1}$.
The above mentioned analysis of the X(3872)$\rightarrow$$J$/$\psi$$\pi^+$$\pi^-$ 
was (in addition to the above mentioned results) able to determine an upper limit
on the width of the X(3872) of $\Gamma$$\leq$1.2~MeV (90\% C.L.). 
The was only possible using a 3-dimensional fit (see Fig.~\ref{fx3872}).
The kinematical over-constraint thus provided access to observables smaller 
than detector resolution of about $\simeq$3$-$4~MeV.
Belle II will be able to perform a width measurement in another decay channel
X(3872)$\rightarrow$$J$/$\psi$$\gamma$, for which the branching fraction is
about one order of magnitude smaller than for $J$/$\psi$$\pi^+$$\pi^-$ .
The expected integrated yield will $\simeq$1750 events, compared to 
151$\pm$15 events for $J$/$\psi$$\pi^+$$\pi^-$ in the total Belle data set. 
The monoenergetic photon in $J$/$\psi$$\gamma$ will provide an additional constraint
and an upper limit of $\leq$1 MeV will be feasible.\\
As another future experiment \panda at FAIR (Facility for Antiproton and Ion
research) at GSI Darmstadt, Germany, will be using cooled antiproton beams.
The cooling will use both stochastic cooling and $e^-$-cooling techniques,
providing a momentum resolution of the antiproton beam of down to
$\Delta$$p$/$p$$\geq$2$\times$10$^{-5}$.
The planned luminosity of {\cal L}=2$\cdot$10$^{32}$~cm$^{-2}$~s$^{-1}$ would translate
into a number of 2$\cdot$10$^9$ $J$/$\psi$ per year, 
if running only at the $\sqrt{s}$=$m$($J$/$\psi$). 
Detailed Monte-Carlo simulation studies of a resonance scan 
for $p$$\overline{p}$$\rightarrow$X(3872) at \panda were performed. 
For 20 scan points with data taking of 2 days each and an input width
$\Gamma_{X(3872)}$ a reconstructed width of 86.9$\pm$16.8~keV
can be achieved. For details see \cite{menu2010} \cite{galuska_master} \cite{galuska_bormio}.

\vspace*{-0.3cm}

\section{Summary}

$e^+$$e^-$ collisions enable unique precision tests of the $q$$\overline{q}$ potential
in the charmonium and bottomonium region. 
The standard potential model fails for many observations, clearly indicating
non-$q$$\overline{q}$ phenomena. Future facilities (Belle II, \panda) will provide 
precision tests not only of masses, but also widths in the sub-MeV regime. 


\end{document}